# Deep UV photolithography enhanced geometric homogeneity for low loss photonic crystal waveguides

*Yahui Xiao[1†], Feifan Wang[1†], Dun Mao[1†], Thomas Kananen[1], Tiantian Li[1], Hwaseob Lee[1], Zi Wang[1], and Tingyi Gu[1*]*

[1] Department of Electrical and Computer Engineering, University of Delaware, Newark, DE 19716, USA

**ABSTRACT.** Periodic or gradient subwavelength structures are basic configurations of photonic crystals and metamaterials. While the numerical simulations predict trivial loss, but the fabricated photonic crystal waveguides by E-beam lithography and standard dry etching shows unacceptable loss beyond 10 dB. Nanofabrication variation introduced geometric inhomogeneity is considered as the primary cause of the deleterious performance. The deep UV photolithography in CMOS foundry is a large-scale parallel processing, which can significantly suppress the fabrication-related inhomogeneous offsets and thus the optical linear loss. Here we demonstrate ultra-low loss photonic crystal waveguides with a 300 mm multi-project wafer run. For sub-millimetre long photonic crystal W1 waveguides, consistent performance of 2 dB total loss and 40 dB extinction ratio are observed across different dies.

## 1. INTRODUCTION

The capability of manipulating photons within ultracompact dimensions makes photonic crystal (PhC) an appealing building block for small footprint and energy-efficient photonic components[1]. PhC line-defect waveguides (WGs) are compact implementation of slow-light optical delay lines[2,3], directional couplers[4,5], and active components with enhanced light-matter interactions[3,6-9]. Nanomanufacturing by the foundry processing dramatically expands the scalability of silicon photonics for technical transitions to industrial applications[10,11]. However, few PhC based structures are found in the process design kits of semiconductor foundries due to the high linear loss of most experimentally demonstrated devices (typically ~ 10 dB per PhC component). The linear loss in PhC WGs is attributed to geometric disorder induced propagation loss from scattering[12] and mode mismatch on the PhC WG coupler[13]. The foundry nanomanufacturing process effectively suppresses the scattering loss in two ways: (1) the deep UV (DUV) photolithography improves the geometric uniformity[14]; (2) the optimized dry etching process reduces side-wall roughness. Currently, the foundry processing optimization has been focusing on PhC high $Q$ cavities[15-18]. Involving those sub-million $Q$ cavities in photonic integrated circuits (PICs) still requires supporting low-loss PhC WGs structures.

In this work, we demonstrate low-loss PhC WGs manufactured by silicon photonics foundry. The total loss, including both insertion loss and the propagation loss over 0.2-mm-long PhC WG, is measured to be around 2 dB in multiple devices across different dies. The extinction ratio between the pass and stop band is measured to be around 40 dB. The minimal tunneling photons in the stopband indicate excellent geometry homogeneity. Those devices are manufactured through a multi-project wafer run on a 300-mm wafer and post-processed in the local cleanroom for the undercut. With the carefully designed undercut profile, the two-stage three-dimensional coupler between channel WGs on buried oxide and suspended PhC membrane shows less than 1 dB insertion loss. The superior device performance (Table I) promises PhC in large-scale multi-components PICs for optical signal processing and interconnects.

Table I: Summary of PhC WGs' losses and extinction ratio on the edge of line defect bandwidth

| Ref. | Lithography | Coupler loss (per port) | Propagation loss | Total PhC loss | Extinction ratio |
| --- | --- | --- | --- | --- | --- |
| (19) | DUV | 8.5 dB | 10 dB/cm | 10 dB | 40 dB |
| (20) | Ebeam | ~ 2.5 dB | -- | ~ 16 dB | 25 dB |
| (13) | Ebeam | 4 dB | -- | 9 dB | 22 dB |
| (21) | DUV | 4 dB | 14.2 dB/cm | 10 dB | -- |
| (22) | Ebeam | -- | -- | 20 dB | 20 dB |
| (24) | Ebeam | 2 dB* | 24 dB/cm | 5 dB | 32 dB |
| (5) | Ebeam | 3-4 dB | 10 dB/cm | 4 dB | 25 dB |
| This work (Fig. 1(a)) | DUV | < 1 dB | <10 dB/cm | ~2 dB | 40 dB |
| UDNF (Fig. 1(b)) | Ebeam | -- | -- | 25 dB | 20 dB |

*polymer coupler

## 2. RESULTS AND DISCUSSION

The geometric homogeneity of the PhC devices is studied by spatial Fourier transform. The scanning electron microscope (SEM) images of the fabricated PhC devices are shown in Fig. 1(a)-(b). Fig. 1(a) is the SEM image of the foundry manufactured PhC WG defined by the DUV photolithography. The identical layout design is fabricated in the university cleanroom as a control sample (Fig. 1(b)). The spatial periodicity of both PhC structures is verified by the bright dots in the spatial frequency domain. The blurred lines are results from the line defect of WG and the cavity. The better homogeneity of the foundry manufactured device is confirmed by (1) extended dots array in two dimensions (Fig. 1(b)) and (2) lower background signal in the spatial frequency domain (the orange curve in Fig. 1(e)).

Fig. 2(a) illustrates the workflow of the foundry manufactured PhC devices. We designed the W1 PhC WG with a targeting wavelength of around 1550 nm. The lattice constant of the PhC air holes is fixed as 440 nm, with the radii vary from 135 nm to 170 nm by an increment of 5 nm for the geometric offsets. After the pattern layout pass the design ruler check (DRC) for foundry compatibility, the PhC devices defined by DUV photolithography are manufactured as an multi-project wafer (MPW) passive run at AIM Photonics. The post-processing by wet etching undercut for the certain PhC WG regions to increase the refractive index contrast. Fig. 2(b) shows the optical image of multiple dies from the MPW run, where the inset is a magnified view of a single die. Arrays of PhC devices are manufactured on a 300 mm silicon-on-insulator (SOI) wafer with 220 nm-thick device layer. The lattice constant and radius of the PhC are more than the critical dimension of the foundry (440 nm and 155 nm, respectively). The silicon layer is protected by a thick PECVD silicon oxide film. Post-processing selectively etches the substrate and superstrate of the PhC area (Fig. 2(c)). Firstly, this top oxide is removed by wet etching with buffered oxide etching (BOE) 1:6 at a rate of about 160 nm/min. A photoresist (AZ1512) is then spin-coated onto the sample, followed by laser writing to open a window above the PhC region. The window shrinks to the inner region to keep the edges of PhCs untouched by the liquid. The zig-zag shape helps to keep the PhC stable, even when it is suspended. After the laser writing, the sample is dipped into BOE 1:6 to do the undercut. The sample is kept in BOE etchant for 30 minutes. Finally, the remaining resist is stripped. Fig. 2(d) shows the optical images of the device near the PhC WG-channel WG interface for steps (i), (iv), and (vi), as shown in Fig. 2(c). After the wet etching, the underlying oxide in the polygon area is removed. After stripping off the resist, the suspended area changes its structural color from pink to green. The green color indicates the thermal oxide under the PhC is removed. Measurements show that the etched area is larger than the defined area about 1 micrometer in the plane. As the etching is isotropic, we estimate 1-micron thick thermal oxide is removed under the PhC plane.

Fig. 3(a) represents the side-view design of the two-stage channel WG – PhC WG coupler, where the input light travels from the low loss channel WG to the supported PhC WG, and from the supported PhC WG to the suspended region. The overall loss is measured to be less than 1 dB. Fig. 3(b) shows the top view of the optical microscope image on the coupling region, where the dashed curve represents the interface between the supported and the suspended region. The green color inside the window indicates the suspended region after undercut. Scanning electron microscope images verify little oxide or resist polymer residue. Fig. 3(c) shows the perspective view of the area near the PhC WG and the edge of the window. The SEM images show little residue and nonuniformity of the periodic PhC structure. The zoomed view of the dashed box at the interface between supported and suspended PhC WG after post-processing is shown in Fig. 3(d). Inside the window defined on the photoresist, the thermal oxide under the PhC layer is removed to reduce light coupling to the substrate. At the same time, there is some residual left outside the window to improve the mechanical stability of the PhC structure.

To characterize the loss of the PhC device, we measured their broadband transmission spectra and compared the results across the dies. A continuous-wave laser tunable from 1480 nm to 1580 nm is directed to the input grating coupler by single-mode fibers. The signal from the output grating coupler is connected to a photodetector. We tested the PhC WG transmission spectra before and after the undercut. The transmission spectra of PhC WG with the hole radius of 140 nm and 135 nm in chiplet I are shown in Fig. 4(a) and (b), respectively. Before the undercut (grey curves), as shown in the (iv) step of the fabrication process, the total loss of the PhC device is larger than 10 dB, and there is no clear band-edge between 1480-1580 nm. After the undercut (orange curves) in the (vi) step, a clear and sharp band-edge appears with an extinction ratio over 40 dB. The flat curve before the band edge indicates an insertion loss around 2 dB for this 0.2-mm-long PhC WG. Similarly, Fig. 4(c) and (d) represent the PhC WG with the working radius of 155 nm and 150 nm in chiplet II, respectively. A significant sharp band-edge appears, and the extinction ratio reaches over 30 dB. The difference in the working radius between chiplet I and chiplet II is the minute post-processing time difference during the wet etching. In this case, chiplet I has a slightly longer time sinking in the BOE solution than chiplet II, which caused the working radius to shift to a smaller size.

**DISCUSSION**

In summary, we experimentally demonstrate low-loss PhC WGs with foundry processing and post-processing. The high extinction ratio near the PhC WG passband edge means the negligible tunneling photons through the defect states in the PhC structure, indicating excellent geometric homogeneity. The scalability and repeatability of the low-loss PhC WGs set a baseline for adopting the PhC based compact devices into the industry-standard photonic foundry toolbox, towards large-scale silicon photonic circuits for optical interconnects and quantum information processing.

**METHODS**

Layout Preparation: The W1 PhC WG is designed by the Finite-difference time-domain method[24], with operation wavelength in telecommunication C band. With a fixed lattice constant of 440 nm, the geometric offsets are introduced into the void radii (from 135 nm to 170 nm with an increment of 5 nm). The width of the channel WGs is fixed to be 500 nm.

Fabrication: Those DUV photolithography defined PhC devices are manufactured as an MPW passive run at AIM Photonics. The PhC device in Fig. 1(b) is fabricated at the university cleanroom. The PhC layout was firstly defined in CSAR 6200.09 positive resist layer by using a Vistec EBPG5200 electron beam lithography system with 100 kV acceleration voltage, followed by resist development and single-step dry etch procedures.

Measurement: We measured the transmission spectrum and insertion loss of the PhC WG through the devices by a coupled fiber system. A continuous-wave laser tunable from 1480 nm to 1580 nm (with a spectral resolution of 10 pm) is directed to the input grating coupler by single-mode fibers. The signal from the output grating coupler is connected to a photodetector.

## ASSOCIATED CONTENT

**Correspondent Author**

Tingyi Gu – Electrical and Computer Engineering, University of Delaware, Delaware, United States; *E-mail: tingyigu@udel.edu

**Author Contributions**

F. W., Y. X., and D. M. performed post-processing and measurements. T. K. and T. L. prepared the TAPEout. F. W., Y. X., D. M., and T. G. analyzed the device results. Y. X., F. W., and T.G. wrote the manuscript, and all authors discussed the results and commented on the manuscript.

†Y. X., F. W., and D. M. contributed equally to this work.

**Notes**

The manufacturing process information is not available according to the non-disclosure agreement between the authors' institution and AIM Photonics.

## ACKNOWLEDGMENTS

**TOC**

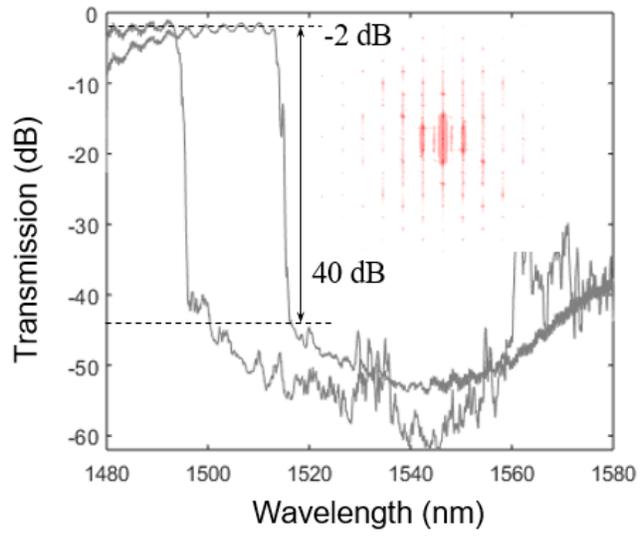

**FIGURES:**

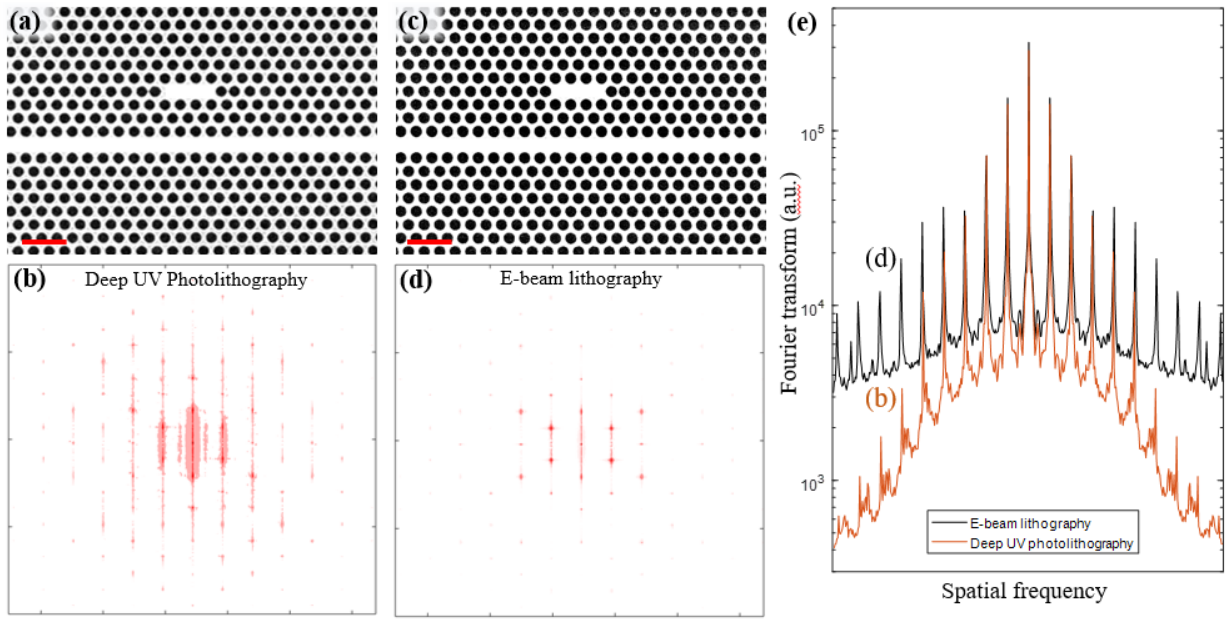

**Figure 1. Comparison of the geometric homogeneity of the photonic crystal (PhC) structures fabricated by DUV photolithography in a silicon photonic foundry and e-beam lithography in the local cleanroom.** (a) Example SEM image of foundry processed PhC structure and (b) correspondent two-dimensional spatial Fourier transform plot in log scale. (c) and (d) similar PhC design fabricated by the e-beam lithography. Scale bars: 1µm. (e) Comparison of the spatial spectra of the images shown in (a) and (b). The reduced background noise (red curve) indicates higher spatial uniformity of the photolithography defined periodic structure.

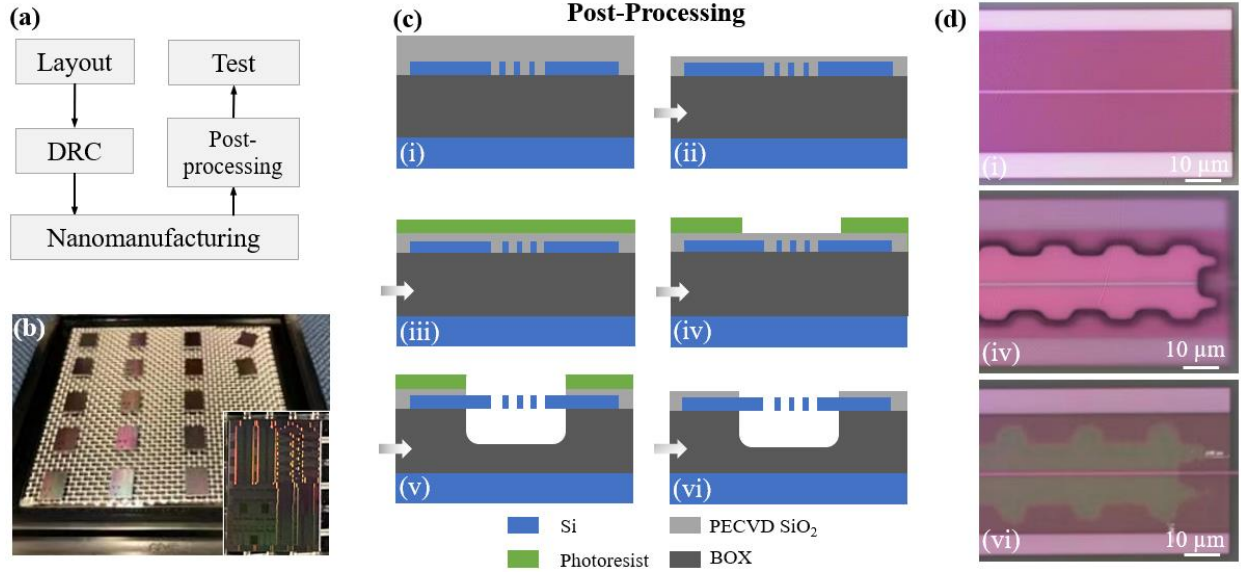

**Figure 2. Preparation of the PhC chiplets.** (a) Workflow of the foundry manufactured PhC devices. DRC: design ruler check for foundry compatibility. (b) Optical images of multiple dies received from the foundry. Inset: Top view of a single die. (c) Postprocessing flow chart: (i) Foundry manufactured Si PhC; (ii) Top oxide removal; (iii) Spin coating photoresist; (iv) Photolithography; (v) Undercut; (vi) Stripping resist. (d) The optical microscope images of the PhC devices at step i (top), iv (middle), and vi (bottom). The undercut region is a polygon with zig-zag edges to improve the mechanical strength of the membrane.

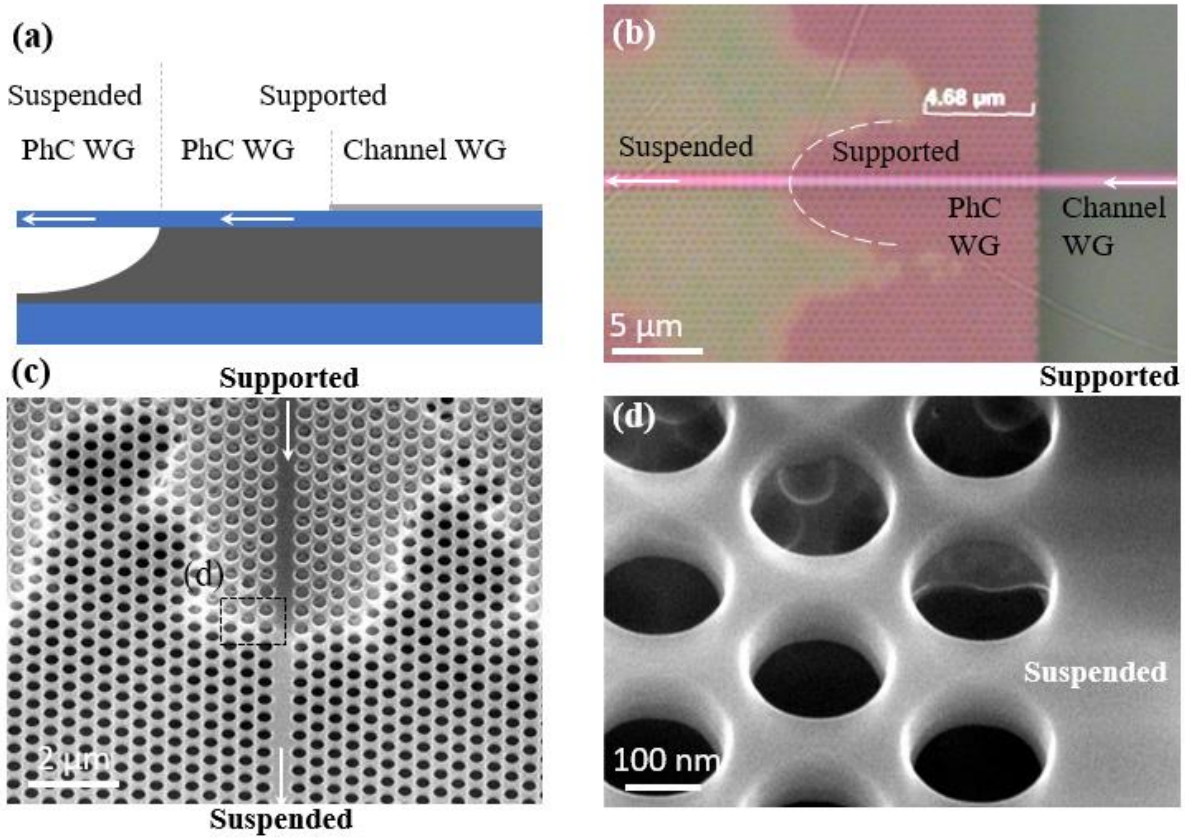

**Figure 3. Low loss three-dimensional coupler between a channel waveguide and a PhC waveguide.** (a) Design of the channel WG – PhC WG coupler. (b) Top view of SEM image and (c) perspective view of the optical image on the coupling region. (d) Close view of the interface between supported and suspended PhC WG after post-processing.

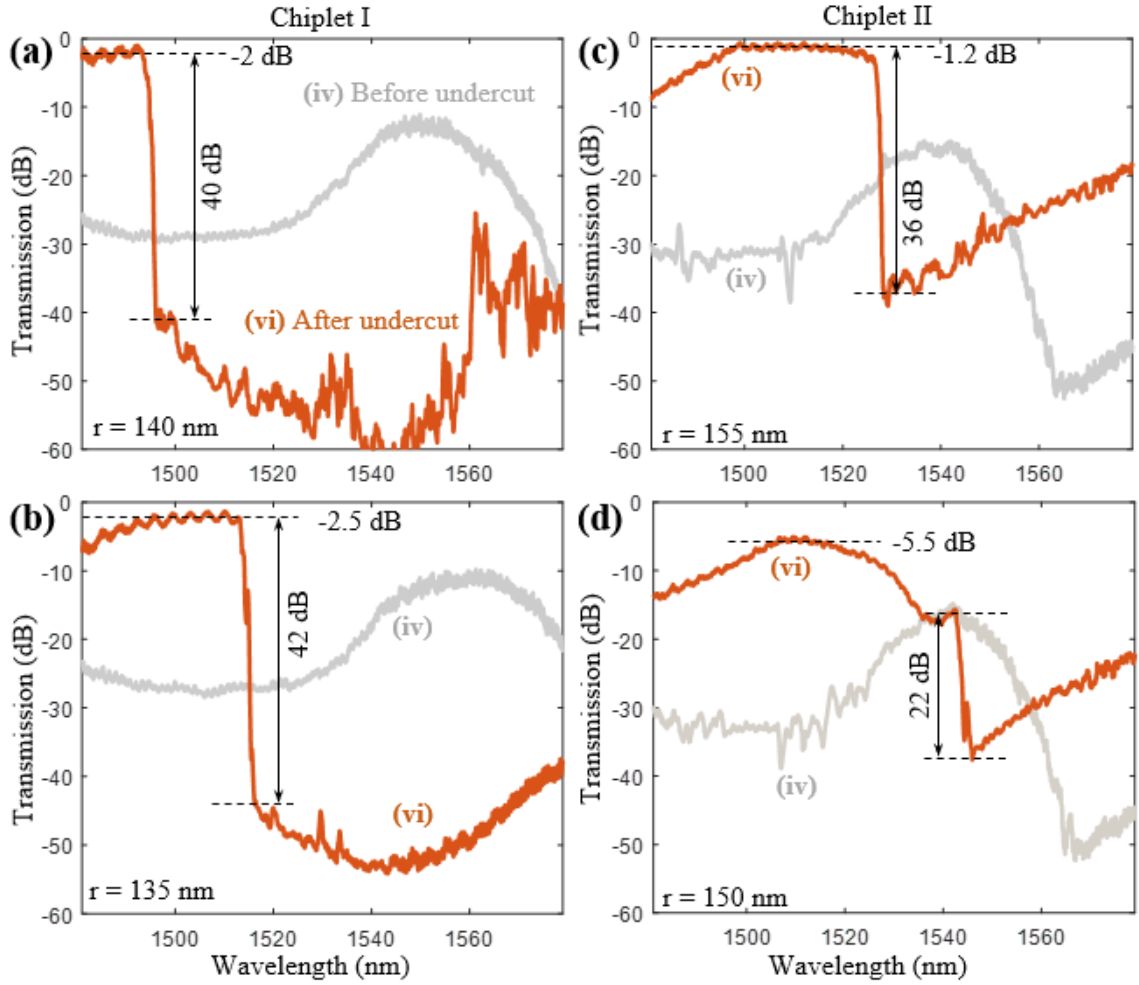

**Figure 4. Measured total loss of manufactured PhC WGs across dies** (grating coupler loss excluded)**.** The PhC lattice constant is at 440 nm and (a) the radius r=140 nm PhC device before (grey) and after undercut (orange); (b) r=135 nm; (c) r=155 nm; (d) r=150 nm.